\documentclass[12pt]{article}

\usepackage{epsfig,amsmath,amssymb}
\usepackage{color}
\usepackage{graphicx}
\usepackage{cite}  
\usepackage{epsfig,amsmath,amssymb}
\usepackage{color}

\newcommand{\be}{\begin{equation}}
\newcommand{\ee}{\end{equation}}
\newcommand{\ra}{\rangle}
\newcommand{\la}{\langle}
\newcommand{\bit}{\begin{itemize}}
\newcommand{\eit}{\end{itemize}}
\newcommand{\bea}{\begin{eqnarray}}
\newcommand{\eea}{\end{eqnarray}}

\newcommand{\Neel}{N\'{e}el}
\newcommand{\bfs}{{\bf s}}

\begin{document}
\title
{The spin-1/2 square-lattice $J_1$-$J_2$ model: The spin-gap issue}

\author
{ J. Richter$^1$, R. Zinke$^2$, and D.J.J. Farnell$^3$\\
\small{$^1$ Institut f\"ur Theoretische Physik, Universit\"at Magdeburg, 39016 Magdeburg, Germany}\\
\small{$^{2}$ Institut f\"ur Apparate- und Umwelttechnik, Universit\"at Magdeburg, 39016 Magdeburg, Germany}\\
\small{$^{3}$ 
   School of Dentistry, Cardiff University
   Cardiff CF14 4XY, Wales UK}
}

\date{\today}

\maketitle

\begin{abstract}
We use the coupled cluster method to high orders of approximation
in order to calculate the ground-state energy, the ground-state  magnetic order parameter, and the spin gap of
the spin-1/2 $J_1$-$J_2$ model on the square
lattice. 
We obtain values
for the  transition points to the magnetically disordered quantum
paramagnetic phase of $J_2^{c1}=0.454J_1$ and  $J_2^{c2}= 0.588 J_1$.
 The spin gap is zero in the entire parameter region accessible by our
 approach, i.e. for $J_2 \le 0.49J_1$ and $J_2 > 0.58J_1$. This finding is in favor of a gapless spin-liquid
 ground-state in this parameter regime.

\end{abstract}


75.45.+j Macroscopic quantum phenomena in magnetic systems


\section{Introduction}
\label{intro}
The spin-1/2 quantum Heisenberg antiferromagnet with nearest-neighbor (NN),
$J_1 > 0$,  
and next-nearest-neighbor (NNN) bonds, $J_2\geq 0$, described by the
Hamiltonian
\be\label{ham}
 H =  J_1\sum_{\langle ij \rangle} \bfs_i \cdot \bfs_j + J_2\sum_{\langle\langle ij
 \rangle \rangle}\bfs_i \cdot \bfs_j
\ee
is one of the most
challenging quantum spin models. The model was introduced more than 25 years
ago in order to describe the breakdown of  \Neel\ antiferromagnetic  (NAF) long-range
order (LRO) in cuprate superconductors.
\cite{inui88,chandra88,dagotto89}

This model has since attracted much attention as a canonical model for
studying frustration driven quantum phase transitions between semiclassical
ground-state phases with magnetic LRO and magnetically disordered quantum
phases, see, e.g., Refs.~\cite{inui88,chandra88,dagotto89,schulz,richter93,richter94,
zhito96,Trumper97,bishop98,singh99,sushkov01,capriotti01,rgm_ihle,Sir:2006,Schm:2006,mambrini2006,darradi08,
sousa,bishop08,xxz,ortiz,singh2009,cirac2009,ED40,fprg,cirac2010,yu2012,balents2012,Mezzacapo,becca2012,verstrate2013,becca2013,beach2013,gong2014,doretto2014,Qi2014,eggert2014,Ren2014,wang2014,chou2014}. 
These studies demonstrate clearly that there is a magnetically disordered
spin-rotation-invariant
quantum
phase in the region of strong frustration $0.44 \lesssim J_2/J_1 \lesssim
0.6$.   
Although the semi-classical ground-state phases of the model, namely
the NAF LRO at 
$J_2/J_1 \lesssim 0.44$
and the  collinear antiferromagnetic  (CAF) LRO  at $J_2/J_1 \gtrsim 0.6 $ are well-understood, 
an active  controversial debate has started very recently regarding the nature of
the intermediate quantum
phase in this model and its quantum phase
transitions.\cite{Sir:2006,darradi08,yu2012,balents2012,Mezzacapo,becca2012,verstrate2013,becca2013,beach2013,gong2014,doretto2014,Qi2014,eggert2014,Ren2014,wang2014,chou2014}
Most of the papers argue that the transition at about  $J_2/J_1 \approx
0.44$
is continuous, but that the transition at about  $J_2/J_1 \approx 0.6$ is of first
order, see, e.g.,
Refs.~\cite{schulz,singh99,sushkov01,yu2012,becca2012,chou2014}.
A particular focus of these  recent papers has been on the existence of an
excitation gap in the intermediate quantum phase.
Those papers in favor of an excitation gap are
given by Refs.~\cite{ED40,balents2012,becca2012,doretto2014,eggert2014}.
We remark that a finite gap  between a spin-rotation-invariant singlet ground state
and a magnetic triplet excitation (spin gap) would be in accordance with earlier results in
favor of a valence-bond ground
state breaking translational symmetry, see e.g.,
Refs.~\cite{dagotto89,richter93,zhito96,singh99,Sir:2006,mambrini2006,darradi08}. 
In contrast to these findings, there are several recent investigations
reporting 
indications of a gapless
spin liquid state.\cite{Mezzacapo,verstrate2013,becca2013,chou2014}
Very recent calculations using density matrix renormalization group
 with explicit implementation of $SU(2)$ spin
rotation symmetry  
in Ref.~\cite{gong2014} have found a gapless spin liquid for $0.44 < J_2/J_1
< 0.5$ and a gapped plaquette valence bond phase for  $0.5 < J_2/J_1 <
0.61$.

In addition to the basic theoretical interest in this frustrated quantum many-body model,
we mention that interest in this model 
is motivated also by its
relation to experimental studies  of various magnetic materials, such as VOMoO$_4$
(Ref.~\cite{VOMoO4}), $\mathrm{Li}_{2}\mathrm{VOSiO}_4$,
and $\mathrm{Li}_{2}\mathrm{VOGeO}_4$ (Ref.~\cite{melzi}).
However, none of these materials has as yet exhibited exchange parameters
$J_1$
and $J_2$ suitable for a magnetically disordered phase at very low
temperatures.

In this paper we focus on the calculation of the gap to 
triplet excitations (spin gap) using a very
general {\it ab initio}  many-body technique, the
coupled-cluster method (CCM), that was successfully applied in various fields of many-body
physics.\cite{ccm_theory}
The spin gap can be calculated directly within the framework of the CCM by
using an appropriate excited-state formalism.\cite{emrich81,bishop00,krueger00,farnell2009}   

\section{Brief illustration of the coupled-cluster method}
\label{ccm}
We illustrate here only some relevant features of
the CCM.
For more general information on the methodology of the CCM, see, e.g.,
Refs.~\cite{ccm_theory,bishop00,zeng98,bishop04,farnell2009}.
The CCM 
has recently been applied widely to frustrated quantum spin systems, see, e.g., 
Refs.~\cite{bishop98,krueger00,ccm_shastry,Schm:2006,darradi08,xxz,ccm_m_h,farnell2009,ccm_j1j2_ferri,richter10,ccm_kagome,honey2011,bishop2013a,bishop2013,bishop2014,jiang2014,
ccm_archimedean}.  
In particular,  the CCM
has been applied to calculate the ground-state properties of model (\ref{ham}) in two recent
publications\cite{darradi08,richter10},  
although the spin gap of the model has never
before been
calculated by using the CCM.

We mention firstly that the CCM automatically yields results in the limit $N\to\infty$.  
Here we follow 
Refs.~\cite{darradi08} and \cite{richter10} to calculate the
ground-state properties of the model using the CCM.
The starting point for a CCM
calculation is the choice of a normalized reference (or model) state
$|\Phi\rangle$. We then define a set of mutually commuting multispin
creation operators $C_I^+$ with respect to this state, where the
index  $I$ runs over a complete set of
many-body configurations.  For the system under consideration here we
choose as 
CCM reference states the two-sublattice N\'{e}el state for small $J_2/J_1$ 
and 
for large $J_2/J_1$ one of two possible
collinear striped states.
Although these CCM reference states are magnetically ordered states,  various 
applications of the CCM to one- and two-dimensional quantum spin systems
demonstrate that  
high-order implementations of the CCM are appropriate to describe magnetically
disordered ground-state phases, see e.g.
Refs.~\cite{krueger00,ccm_shastry,Schm:2006,darradi08,xxz,ccm_kagome,bishop2013a,bishop2013,bishop2014,jiang2014,
ccm_archimedean}.

We perform a rotation of the local axis of
the spins such that all spins in the reference state align along the
negative $z$ axis.  In the rotated coordinate frame the reference state
reads \\
$|\Phi\rangle \hspace{-3pt} = \hspace{-3pt}
|\hspace{-3pt}\downarrow\rangle |\hspace{-3pt}\downarrow\rangle
|\hspace{-3pt}\downarrow\rangle \ldots \,$, and we can treat each site
equivalently.  The corresponding multispin creation operators are written as $C_I^+=s_i^+,\,\,s_i^+s_{j}^+,\,\,
s_i^+s_{j}^+s_{k}^+,\cdots$, where the indices $i,j,k,\dots$ denote
arbitrary lattice sites.

The CCM parameterizations of the ket- and bra- ground states are given
by
\begin{eqnarray}
\label{ket}
H|\Psi\rangle = E|\Psi\rangle ; 
\qquad  
\langle\tilde{\Psi}|H = E\langle\tilde{\Psi}| ;
\nonumber\\
|\Psi\rangle = e^S|\Phi\rangle, 
\qquad 
S = \sum_{I \neq 0}{\cal S}_IC_I^+ ; 
\nonumber\\
\langle\tilde{\Psi}| =  \langle\Phi|\tilde{S}e^{-S},
\qquad 
\tilde{S} = 1 + \sum_{I \neq 0}\tilde{\cal S}_IC_I^- .
\end{eqnarray}
By using the Schr\"odinger equation, $H|\Psi\ra=E|\Psi\ra$,
we can write the ground-state energy as $E_{\rm GS}=\la\Phi|e^{-S}He^S|\Phi\ra$.  The
magnetic order parameter (sublattice magnetization) is given by
\be 
\label{mag}
m_s = -\frac{1}{N} \sum_{i=1}^N \la\tilde\Psi|s_i^z|\Psi\ra ,
\ee
where $s_i^z$ is expressed in the rotated coordinate system. 
The ket-state and
bra-state correlation coefficients are obtained by solving the CCM
ket- and bra-state equations given by
\begin{eqnarray}
\label{ket_eq}
\langle\Phi|C_I^-e^{-S}He^S|\Phi\rangle = 0,
\qquad 
\forall I\neq 0,
\\
\label{bra_eq}
\langle\Phi|\tilde{\cal S}e^{-S}[H, C_I^+]e^S|\Phi\rangle = 0,
\qquad 
\forall I\neq 0.
\end{eqnarray}
Each ket- or bra-state equation belongs to a certain creation operator
$C_I^+=s_i^+,\,\,s_i^+s_{j}^+,\,\, s_i^+s_{j}^+s_{k}^+,\cdots$,
i.e. it corresponds to a certain set (configuration) of lattice sites
$i,j,k,\dots\;$. For the problem at hand only those correlation coefficients ${\cal S}_I$ and  $\tilde{\cal S}_I$   
related to clusters with even numbers of spin flips are different from zero.
The
ket or bra ground states belong to total spin $S^z=0$. 
We use an enlarged unit cell of four sites for the CCM
calculations in order to allow for the possibility of CCM
solutions that break translational symmetry, e.g., for gapped
valence-bond ground states.

We use the well established LSUB$n$ approximation scheme
in order to truncate the expansion of $S$ and $\tilde {S}$, cf., e.g.,
Refs.~\cite{bishop00,bishop04,darradi08,richter10,ccm_kagome,bishop2013a,bishop2013,bishop2014,jiang2014,
ccm_archimedean}.
Within the  LSUB$n$ scheme 
all multispin correlations  over all distinct locales on the lattice
defined by $n$ or fewer contiguous sites are taken into account in the correlation operators $S$ and 
$\tilde {S}$. 
Although the ground state is not in the focus of the present paper, we present here results including the LSUB12 
approximation, which goes 
beyond the LSUB10 approximation presented  in Refs.~\cite{darradi08} and
\cite{richter10}. This increase in the level of approximation yields an improvement of
the accuracy of the ground-state data.

In order to calculate the gap to triplet excitations we follow Refs.~\cite{krueger00}
and \cite{farnell2009},  where the
spin gap was calculated by CCM for the two-dimensional $J-J'$ model\cite{J_Jprime}
and the one-dimensional $J_1$-$J_2$ model,\cite{mikeska} respectively.
Both models exhibit a quantum phase transition from a gapless phase 
to a gapped valence-bond phase. It was found in Refs.~\cite{krueger00}
and \cite{farnell2009}
that the opening of the spin gap at
the transition point to the valence-bond phase is well described by the CCM.
     
To obtain the excited state $|\Psi_e\ra$ from the ground state $|\Psi\rangle$ (\ref{ket})
we apply 
an excitation operator $X^e$ linearly to  $|\Psi\rangle$,
such that 
\be \label{ket_ex} |\Psi_e\ra=X^ee^S|\Phi\ra; \quad X^e=\sum_{I\neq 0}{\cal X}_I^eC_I^+ .\ee
Using the Schr\"odinger equation, $H|\Psi_e\ra=E_e|\Psi_e\ra$, we
find that
\be \label{ket_exc} \Delta_eX^e|\Phi\ra=e^{-S}[H,X^e]_-e^S|\Phi\ra ,\ee
where $\Delta=E_e-E_{\rm GS}$ is the spin gap.
Applying $\la\Phi|C_I$ to Eq.~(\ref{ket_exc}) we
find 
\be \label{ew} \Delta_e{\cal X}_I^e=\la\Phi|C_Ie^{-S}[H,X^e]_-e^S|\Phi\ra ,\ee
which we solve in order to get $\Delta_e$.
The choice of configurations in the excitation operator is restricted   
to contain only those which change the total spin $S^z$ by one.
Hence, the choice of clusters for the excited-state is different
from those for the ground state.   
For the excited state we use the same approximation scheme, LSUB$n$, as for the ground
state thus achieving comparable accuracy for both the ground
and the excited states.
We find that for high orders of approximation the number of configurations
for the excited state is
larger than that for the ground state, i.e. the calculation of
the excited state is more difficult computationally.

The LSUB$n$ approximation becomes exact for $n \to \infty$, and so we
can improve our results by extrapolating the ``raw'' LSUB$n$ data to
$n \to \infty$.  There are well-tested extrapolation
rules\cite{krueger00,ccm_shastry,Schm:2006,darradi08,xxz,farnell2009,ccm_m_h,richter10,ccm_kagome,bishop2013a,bishop2013,
bishop2014,jiang2014,
ccm_archimedean} for the ground-state energy per spin $e_{\rm GS}=E_{\rm GS}(n)/N$,
the magnetic order parameter
$m_s(n)$, and the spin gap $\Delta_e(n)$.
We use $e_{\rm GS}(n) = a_0 + a_1(1/n)^2 + a_2(1/n)^4$ for
the ground-state energy,  
$m_s(n)=b_0+b_1(1/n)^{1/2}+b_2(1/n)^{3/2}$ for the magnetic order parameter,
and 
$\Delta_{e}(n) = c_0 + c_1(1/n) + c_2(1/n)^2$ for the spin gap.
Moreover, we know from Refs.~\cite{bishop08,xxz,darradi08,richter10} that the lowest level
of approximation, LSUB2, conforms poorly to these
rules. Hence, as in previous calculations,
\cite{bishop08,xxz,darradi08,richter10} we exclude LSUB2 data from the extrapolations.

\begin{figure}[ht]
\begin{center}
\epsfig{file=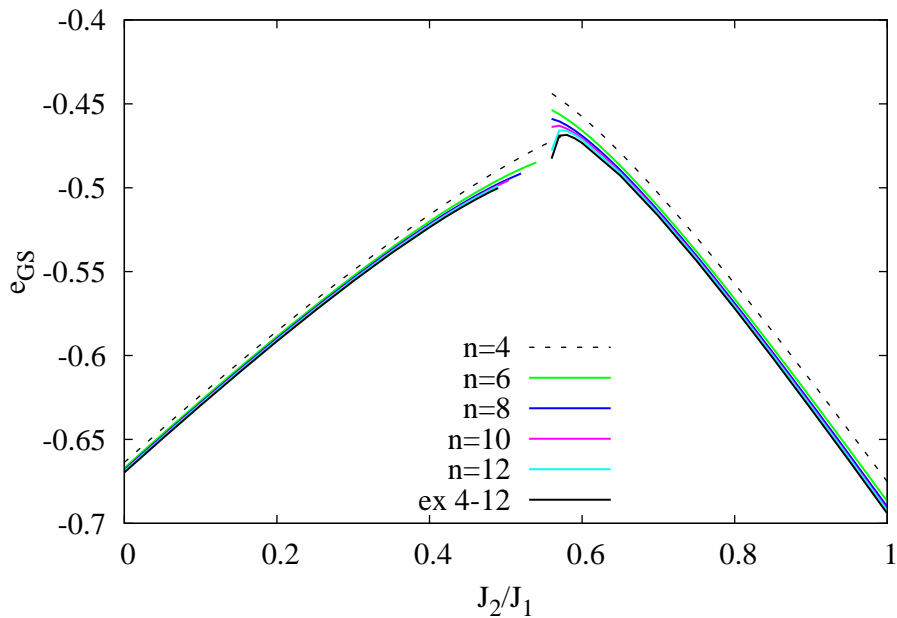,scale=1.0,angle=0.0}
\end{center}
\caption{Ground-state energy per spin plotted as a function of the frustration
parameter $J_2/J_1$. 
}
\label{fig1}
\end{figure}

\begin{figure}[ht]
\begin{center}
\epsfig{file=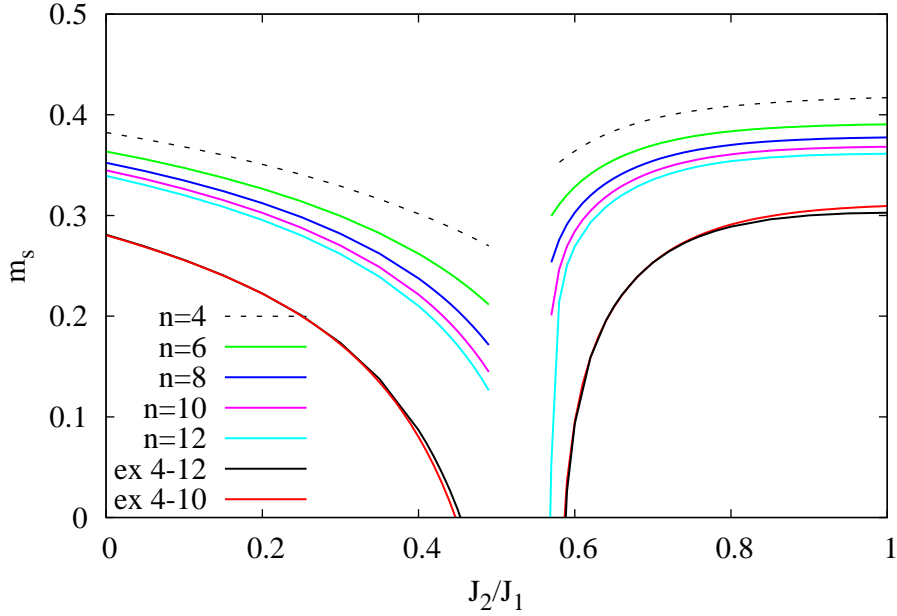,scale=1.0,angle=0.0}
\end{center}
\caption{Magnetic order parameter (sublattice magnetization) $m_s$ plotted
as a function of the frustration
parameter $J_2/J_1$.  
}
\label{fig2}
\end{figure}

\begin{figure}[ht]
\begin{center}
\epsfig{file=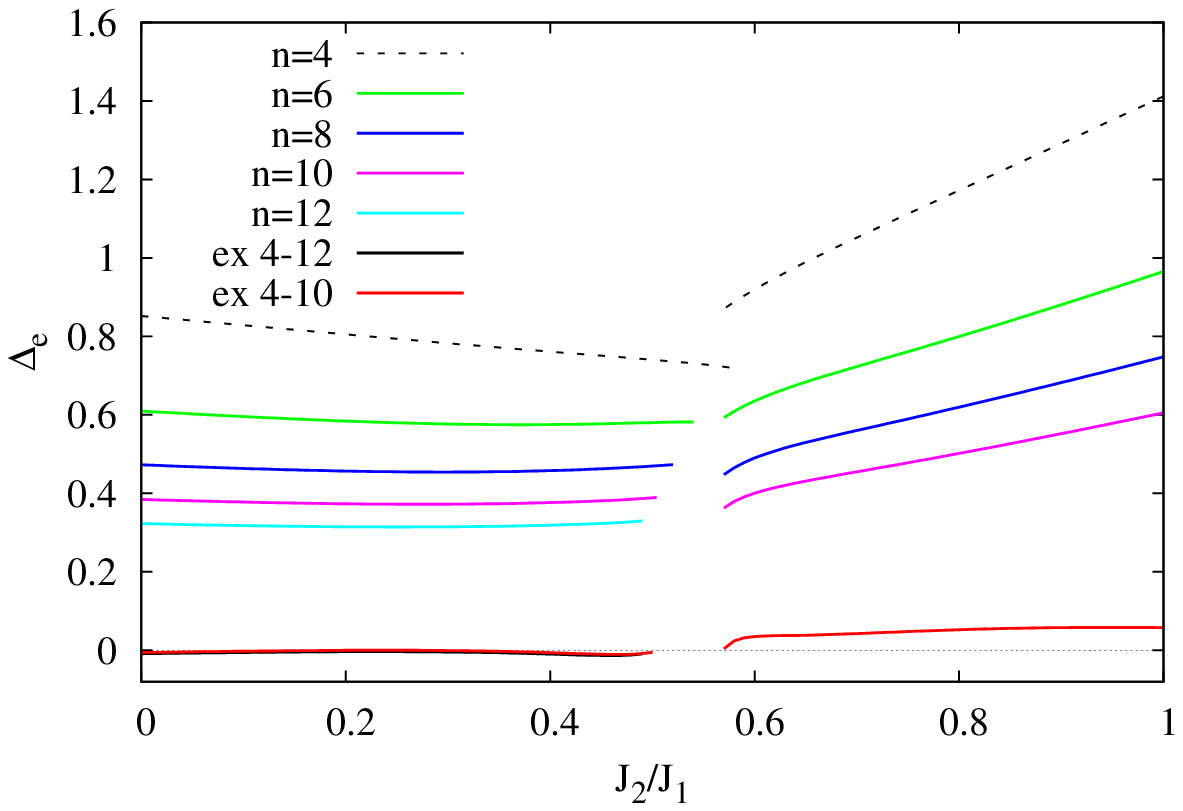,scale=1.0,angle=0.0}
\end{center}
\caption{Spin gap $\Delta_e$
plotted as a function of the frustration
parameter $J_2/J_1$.  Note the beyond $J_2/J_1 =0.58$, where  
the collinear striped state is used as the CCM     
reference state, we are not able to present LSUB12 data, since  
the number of equations which have to be solved 
within the excited-state formalism exceeds the available
computational resources, see also main text. 
}
\label{fig3}
\end{figure}

\section{Results and discussion}

In order to obtain results for the ground-state energy per spin, $e_{\rm GS}$,
the magnetic order parameter (sublattice magnetization),
$m_s$, and the spin gap, $\Delta_e$, we  solved numerically up to
$1,374,389$ equations (for the LSUB12 approximation of the excited state starting with the \Neel\
reference state) using the code of Schulenburg and Farnell.\cite{cccm}
Unfortunately, for the collinear striped    
reference state we can present LSUB12 data for the ground state only. For the spin gap
we are limited to LSUB10, since 
the number of equations is higher
(namely  $2,266,307$ for LSUB12 approximation of the excited state) than
that for the \Neel\ reference state.
We find stable solutions for the CCM equations  at given level of LSUB$n$
approximation for the pure Heisenberg antiferromagnet ($J_2 = 0$) and then we
track this stable solution until it terminates at 
$J_2/J_1=0.59$ (LSUB4), $0.54$ (LSUB6), $0.52$ (LSUB8), $0.50$ (LSUB10),
$0.49$ (LSUB12), respectively.
We notice that this allows to us consider a parameter region that is
noticeably beyond the presumably continuous transition between the NAF phase
and the quantum phase.
On the other side of the phase diagram the transition to the quantum phase is likely first order,
and, therefore, below the CCM transition point at $J^{c2}_2/J_1 = 0.588$
 (see below)
the validity of the CCM data starting from the
collinear striped reference state becomes questionable.

Although our paper focusses primarily
on the calculation of the spin gap, we present firstly
results for the ground-state energy per spin (Fig.~\ref{fig1}) and the
sublattice magnetization (Fig.~\ref{fig2}). The results for the spin
gap are then shown (Fig.~\ref{fig3}).
In all figures we show LSUB$n$ data for $n=4,6,8,10,12$ as well as data
extrapolated to $n=\infty$, where 
extrapolations including new LSUB12 data  are labeled by 'ex 4-12' and the
extrapolations without
including LSUB12 data are labeled by 'ex 4-10'.
Note that the graphs for $e_{\rm GS}$ and
$m_s$ are very similar to those of Ref.~\cite{darradi08}, although  LSUB12 data are now included for
$J_2/J_1 \le 0.49$ and $J_2/J_1 \ge 0.56$.
The comparison of both curves ('ex 4-12' and ''ex 4-10', with and without
LSUB12, respectively) allows us to form an impression on the
accuracy of the
extrapolated CCM results.

Taking into account the new LSUB12 data we find for the phase transition
points
between the semiclassical phases and the quantum 
phase $J_2^{c1}=0.454J_1$ and $J_2^{c2}=0.588J_1$, whereas the
previous CCM values (without LSUB12 data) were $J_2^{c1}=0.447J_1$ and
$J_2^{c2}=0.586J_1$, (see
Ref.~\cite{darradi08}).
These results  are in good agreement with the density matrix
renormalization group data of Ref.~\cite{gong2014}.
Let us mention that many of the earlier attempts to determine the transition
points obtained values $J_2^{c1} \approx 0.4$ (or even smaller values), see
e.g.,
Refs.~\cite{schulz,singh99,rgm_ihle,capriotti01,sushkov01,sousa}.
In view of our results at hand and other recent
results\cite{gong2014,chou2014} the transition point is rather at
higher values $J_2^{c1} \sim (0.44 \ldots 0.45) J_1$.

The ground-state energy $e_{\rm GS}$ (Fig.~\ref{fig1}) shows a monotonic increase with
increasing $J_2$ for the \Neel\ reference state. 
This figure shows also that there is a monotonic increase of $e_{\rm GS}$   with
decreasing $J_2$ (until about
$J_2 \sim 0.57 \ldots 58J_1$,  i.e. slightly
beyond the transition point $J^{c2}_2$), starting from the regime of large $J_2$ and using the collinear
striped
reference state.
For values of $J_2 \lesssim 0.58J_1$, we see that there is a drastic downturn in
$e_{\rm GS}$,
which indicates that the CCM using the
collinear striped reference state becomes inappropriate beyond $J^{c2}_2$.

The results for the magnetic order parameter $m_s$ shown in Fig.~\ref{fig2}  may be used
in order to determine the above reported transition points,  $J^{c1}_2$  and
$J^{c2}_2$, at which the order parameter
vanishes.
Fig.~\ref{fig2} shows continuous behavior for the order parameter near 
 $J^{c1}_2$  and near
$J^{c2}_2$. However, it is clear from this figure that the decrease of the CAF
order parameter to zero at $J^{c2}_2$ is much steeper and more abrupt than the
corresponding decay of the NAF order parameter at $J^{c1}_2$.
This behavior might be an indication  of a continuous transition at  $J^{c1}_2$ and of a first-order transition 
at  $J^{c2}_2$.

Results for the spin gap
$\Delta_e$ are presented in Fig.~\ref{fig3}.  
These data show a monotonic behavior of $\Delta_e$
with increasing level $n$ of the CCM-LSUB$n$ approximation for all values of
$J_2$, which indicates
that the extrapolation to $n \to \infty$  as described  Sec.~\ref{ccm} 
is reasonable. A consistency check of our spin-gap data is
also provided by 
our results
within the semiclassical ground state phases with magnetic
LRO, i.e. for for $J_2 \le J^{c1}_2$   and
$J_2 \ge J^{c2}_2$, where the spin gap has to be zero.
Indeed, our extrapolated results for $\Delta_e$ 
in the limit $n \to  \infty$  are very close to zero within the NAF phase.
In the the CAF phase we obtain a  small finite  $\Delta_e(n=\infty)$, which  indicates that
extrapolations in this regime are less accurate.\cite{comment}

The most relevant and important results for the current discussion relating
the ``spin-gap issue'' concern our results for the spin gap in the
region $J_2 > J^{c1}_2$. 
As already mentioned above we compare two extrapolations
labeled by 'ex 4-12' (including the LSUB12 data in the extrapolation) and by 'ex
4-10'  (without including the LSUB12 data in the extrapolation).
We can consider the difference between both extrapolations as a measure of
the accuracy of the extrapolated spin-gap data. As for the ground-state energy and
the magnetic order parameter both extrapolations for the spin gap almost coincide which
yields evidence that the extrapolation works very well
and our extrapolated data are almost converged, and likely higher-order
approximations would  not modify our results noticeably.
We remark again that the N\'eel reference  state allows us to find
results up to $J_2 = 0.49J_1$ for approximation level LSUB12 (and even larger values
of $J_2$ for lower
approximation levels, see Fig.~\ref{fig3}), which is clearly inside the
magnetically disordered
quantum regime.
We find that there is no significant increase of the LSUB$n$ values for the
spin gap in the parameter region accessible by using the
\Neel\ 
reference state for $J_2 > J^{c1}_2$, i.e. for a considerable region inside the
magnetically disordered quantum phase.
Also the
extrapolated spin gap $\Delta_e(n \to \infty)$ available until $J_2=0.49J_1$ remains practically
zero.
Thus, our CCM results for the spin gap are in favor of a gapless
spin liquid\cite{spinliquid} in accordance with
Refs.~\cite{verstrate2013,becca2013,gong2014,wang2014,chou2014}.
However, our results do not rule out the possibility that a spin gap occurs
within the parameter region  $0.49
\le J_2/J_1 \le 0.59$, as indicated by recent 
density matrix renormalization group calculations.\cite{gong2014} 

Let us briefly discuss these findings in relation to previous CCM
results concerning a possible valence-bond solid ground
state.\cite{darradi08} 
In Ref.~\cite{darradi08} generalized
susceptibilities related to possible valence-bond states were calculated.
Starting from the pure Heisenberg model at $J_2=0$ (i.e., in the  \Neel\  phase),
both susceptibilities for the columnar dimerized and plaquette valence-bond
states grow monotonically with increasing $J_2$, cf. Figs. 8 and 9 in
Ref.~\cite{darradi08}.
These susceptibilities     
become
very large,  but remain finite in the region around 
$J^{c1}_2$;
this behavior was interpreted as an indication for the emergence of a
valence-bond solid phase.
In view of our new CCM results for
the spin gap a more consistent interpretation is that of
enhanced dimer-dimer or/and plaquette-plaquette correlations
which may be even critical (i.e. with a power-law
decay without valence-bond LRO\cite{spinliquid}) in a small but finite regime $0.454 \le J_2/J_1
\le 0.49$, cf. 
also Refs.~\cite{gong2014,wang2014}.
Note, however, that the results for the generalized
susceptibilities presented in Ref.~\cite{darradi08} are 
in accordance with a plaquette ordered quantum phase proposed in
Ref.~\cite{gong2014} for  $J_2/J_1 \gtrsim
0.5$.

\section{Summary}

We have used a general {\it ab initio} many-body technique called the
coupled-cluster method (CCM) to high orders of approximation in order to
calculate ground-state properties and
the triplet excitation gap of the square-lattice 
$s=1/2$ $J_1$-$J_2$ model.
Our results for the transition points, $J_2^{c1}=0.454J_1$ and  $J_2^{c2}= 0.588
J_1$, 
 between the 
semiclassical
ground-state phases with magnetic LRO and the intermediate magnetically disordered quantum
phase are in very good agreement with recent density matrix renormalization
group calculations.\cite{gong2014}
The direct calculation  of the spin gap within the intermediate 
quantum
phase, which is possible until  $J_2/J_1 = 0.49$,  does not give any hint for an
opening of a spin gap in this phase.
Therefore, our results are in favor of a gapless spin-liquid quantum ground state in the region $J_2^{c1} < J_2 \lesssim 
0.49J_1$.
However, a gapped phase for $ J_2/J_1 \gtrsim 0.5$ cannot be ruled out,
because our CCM approach becomes inappropriate for $0.49 \lesssim J_2/J_1
\lesssim 0.58$.    

\hspace{0.5cm}

\section*{Acknowledgments}
For the numerical calculation we used the program package `The
crystallographic CCM' (D. J. J. Farnell and J. Schulenburg).\cite{cccm}



\begin{thebibliography}{999}
\bibitem{comment} The reason for lower accuracy of the spin gap using the collinear striped
reference state  might by  attributed
to the facts that (i) we do not have  LSUB12 data for the extrapolation 
and (ii) $J_1$ bonds within a certain LSUB$n$ approximation belong
to contiguous sites, but  $J_2$ bonds do not (this difference becomes
relevant for larger $J_2$, i.e. in the CAF phase).
\end{thebibliography}
\end{document}